   \documentstyle[aps,twocolumn,prb,psfig]{revtex}

\begin{document}
\draft

\title{Onsager reaction-field theory for magnetic models on 
diamond and hcp lattices}
\author{G.\ M.\  Wysin}
\address{Department of Physics, 
Kansas State University, 
Manhattan, KS 66506-2601 }
\date{September 15, 1999}
\maketitle

\begin{abstract} 
The Onsager reaction-field (ORF) theory is extended so that it applies
to any three-dimensional Bravais lattice with a basis.
The ORF calculation is used to predict the critical temperature
for classical Ising, XY, and Heisenberg magnetic models, in particular, 
on diamond and hexagonal close packed lattices.
Results are compared with series extrapolations and other theoretical
approaches where available.
For the hcp lattice the ORF calculation is seen to be exactly 
equivalent to a Green's function approach.
\end{abstract} 
\pacs{PACS numbers: 75.10.Hk, 75.10.-b, 75.40.Cx}

\section{Introduction}
The Onsager reaction field (ORF) theory\cite{Onsager36} is an improved
form of mean-field theory that includes at least partially
the effects of correlations between nearby atoms.
It was originally applied in magnetism by Brout and Thomas,\cite{Brout67}
and more recently to spin glasses,\cite{spinglass} itinerant electron 
systems,\cite{electron} Hubbard models,\cite{hubbard} and anisotropic 
Heisenberg\cite{heisen} and XY\cite{XY} models.
The procedure is versatile and has been used to estimate specific 
heat, susceptibility and correlations above the critical
temperature, $T_c$, as well as $T_c$ itself.
A clear review of the ORF method applied to three-dimensional
(3D) Ising models is given by White.\cite{White83} 
The method has been applied on the standard Bravais 
lattices, including simple cubic (sc), 
body-centered-cubic (bcc), and face-centered-cubic (fcc),
with results given in terms of integrals over the
associated Brillouin zone (BZ). 
However, a modification of these calculations is needed
to consider other lattices which are not in the Bravais
classification.
Here we show how to apply the ORF procedure to any
non-Bravais lattice that can be considered as an underlying
Bravais lattice with a basis.
In particular, the diamond and hexagonal close packed
lattices are analyzed, both of which have two-atom bases.

In the usual mean-field theory due to Weiss,\cite{Weiss48} 
a chosen atom (or spin, for the magnetic problems we consider) is 
viewed as interacting with the average, or mean-field, of its 
nearest neighbors.
The exact Hamiltonian is replaced by the mean-field one,
in which the neighbors are introduced as the mean-field
acting on the central atom.
However, the central atom itself influences the neighbors,
and therefore the mean-field usually includes a part directly
attributed to the central atom.
This means the mean-field includes a part that might be
considered a self-interaction effect, which should really be
subtracted out.
This results in an overestimate of the critical temperature,
$T_c$.
The ORF procedure is simply a way to estimate and subtract out 
this self-interaction part, i.e., by adding a ``reaction field''
term that accomplishes this.
In this way, the estimate of $T_c$ is brought down, indeed,
usually ORF leads to an underestimate of $T_c$.

The standard ORF approach uses as input the specific lattice 
structure, be it sc, fcc, bcc, etc.  
However, the usual approach and well-known formulas require 
the perfect periodicity of a Bravais lattice--all atoms
are taken as equivalent.
On the other hand, we have been interested in mean-field
and other calculations\cite{Wysin99} for diamond lattices 
because of the low coordination number ($z=4$). 
The diamond lattice is not a Bravais lattice--all sites
do not have the same surroundings, instead, the diamond lattice 
can be considered to be a fcc lattice with a two atom basis.
Thus it is interesting to understand how to apply the ORF
procedure to such a system.
Recently there is interest in ferromagnetic ordering of 
hcp $\ ^3$He at low temperatures,\cite{He3} assumed to be described
by a Heisenberg model.
The hcp lattice is another example of a non-Bravais
lattice; it can be considered as simple hexagonal (stacked
triangular nets) also with a two-atom basis.
Here we show how to extend the standard ORF calculation
of $T_c$ to these two systems, however, our approach
will apply to any Bravais lattice with a basis, i.e.,
any system with multiple atoms per unit cell.

For simplicity we display formulas for Ising models on a 
3D lattice with spins $S_{n}=\pm S$ and coordination number $z$.
However, the modifications to treat $n$-component spins
[i.e., XY (n=2), Heisenberg (n=3), etc. ]
are minimal and will be noted where they are appropriate.
The Hamiltonian is
\begin{equation}
\label{Hamil}
{\cal H}_0 = 
- \frac{1}{2} \sum_{\bf n} \sum_{\bf m} J_{\bf n,m} S_{\bf n} S_{\bf m}
- \sum_{\bf n} H_{\bf n} S_{\bf n}
\end{equation}
where each sum is over all of the lattice sites, 
and the bond coupling strength $J_{\bf n,m}$ depends only
on the neighbor displacement, ${\bf n-m}$, and is of
the same strength $J$ for all near neighbor pairs.
$H_{\bf n}$ is a spatially varying applied field.
%

\section{Onsager Reaction Field Correction: Bravais Lattices}
\label{Bravais}
In the ORF calculation (See Ref.\ \onlinecite{White83} for 
more details), a spin at a chosen site
interacts with the mean-field reduced by a ``reaction field''
that depends on the spin at that site.\cite{Onsager36}
For completeness we summarize key aspects of this
calculation here to see how the extension to a
non-Bravais lattice is accomplished.
%

To effect the reaction term, in the real space Hamiltonian  
an extra self-interaction term is added: 
\begin{equation}
\label{react}
{\cal H}_{\rm rf} = \lambda \sum_{\bf n} S_{\bf n} S_{\bf n}
\end{equation}
This is equivalently a delta-function exchange term of strength
$\lambda$.
The constant $\lambda$ is the reaction-field, which is
determined self-consistently in the calculation, by a 
constraint on the magnetic susceptibility, below.
%

The Hamiltonian ${\cal H}={\cal H}_0+{\cal H}_{\rm rf}$,
transformed into wavevector space is,
\begin{equation}
\label{Hamil-q}
{\cal H} = -{1\over 2} \sum_{\bf q} \left[ 
S_{\bf -q} (J_{\bf q}-\lambda) S_{\bf q}
+ \left( H_{\bf -q}S_{\bf q} + H_{\bf q}S_{\bf -q} \right) \right]
\end{equation}
where the Fourier-space quantities derive from
\begin{equation}
S_{\bf n} = {1\over \sqrt{N}}
   \sum_{\bf q} S_{\bf q} e^{i{\bf q}\cdot{\bf n}}.
\end{equation}
\begin{equation}
J_{\bf n,m} = {1\over N}
   \sum_{\bf q} J_{\bf q}e^{i{\bf q}\cdot({\bf n-m})},
\end{equation}
together with a similar definition for the field $H_{\bf q}=H_{\bf -q}$,
for real applied field $H_{\bf n}$.
$N$ is the number of lattice sites.
The sum in Eq.\ \ref{Hamil-q} is over all ${\bf q}$ in the
appropriate Brilluoin zone.
The Fourier-transformed exchange interaction is
\begin{equation}
\label{Jq}
J_{\bf q}= \sum_{\bf r} J_{\bf r} e^{i {\bf q}\cdot {\bf r}},
\end{equation}
i.e., a sum over displacements to nearest neighbors, 
${\bf r}\equiv {\bf m-n}$.
%

The q-dependent magnetization and zero-field susceptibility have
the usual definitions,
\begin{mathletters}
\begin{equation}
\label{magn}
M_{\bf q}=\langle S_{\bf q} \rangle,
\end{equation}
\begin{equation}
\label{susc}
\chi_{\bf q} =  { \partial M_{\bf q} \over \partial H_{\bf q} }
= \beta \langle S_{\bf -q} S_{\bf q} \rangle.
\end{equation}
\end{mathletters}
Then for an arbitrary Hamiltonian, there results the constraint
(used to determine the reaction field $\lambda$)
\begin{equation}
\label{constraint}
{1 \over N} \sum_{\bf q} \chi_{\bf q} = {\beta \over n} S^{2} \equiv \chi_{0}
\end{equation}
The number of spin components $n$ enters here, in the expression
for $\chi_{0}$, when models other than the Ising model are
considered.

Now we consider the mean-field approximation in q-space, i.e.,
using the random phase approximation (RPA), magnetization components
at different wavevectors are assumed to be independent.
The mean-field Hamiltonian for the interaction of a negative
q-component can be written
\begin{equation}
{\cal H} = - \sum_{\bf q} H^{\rm eff}_{\bf q} S_{\bf -q}
\end{equation}
where the effective (or mean-field) magnetic field is
\begin{equation}
H^{\rm eff}_{\bf q} = H_{\bf q} 
+ \left( J_{\bf q}-\lambda \right) \langle S_{\bf q} \rangle
\end{equation}
From the definition (\ref{susc}), and the relation,
\begin{equation}
\chi_{\bf q} = \frac{\partial M_{\bf q}}{\partial H_{\bf q}^{\rm eff}}
\frac{\partial H_{\bf q}^{\rm eff}}{\partial H_{\bf q}}
\end{equation}
the mean-field Hamiltonian gives the well-known expression,
\begin{equation}
\chi_{\bf q} = {\chi_{0} \over {1-\chi_{0}(J_{\bf q}-\lambda)} }
\end{equation}
The reaction field $\lambda$ is determined by forcing this
expression for $\chi_{\bf q}$ to satisfy the constraint 
(\ref{constraint}).
The critical temperature $T_c$ is determined as the 
temperature at which $\chi_{\bf q=0}$ diverges. 
For lower temperatures the ORF calculation gives a negative 
susceptibility at ${\bf q=0}$, signifying the presence
of the ordered state.
In a continuum limit of the constraint (\ref{constraint}), a short 
manipulation leads to an expression for the critical temperature,
\begin{equation}
\label{Tc-ORF}
k_B T_c = \frac{J_{\bf 0}S^2}{nI} 
\end{equation}
The constant $I$ is given from a q-space integral over
the appropriate Brilluoin zone:
\begin{equation}
\label{ORF-I}
I = \frac{V}{N} \int_{\rm BZ} ~\frac{d^3 q}{(2\pi)^3} ~ 
\frac{J_{\bf 0}}{J_{\bf 0}-J_{\bf q} } 
\end{equation}
and $V/N$ is the specific volume per lattice site
(for example, $V/N = a^3, \frac{1}{2}a^3, \frac{1}{4}a^3$ for sc, 
bcc and fcc lattices, respectively, where $a^3$ is the cubic unit 
cell volume), 
At ${\bf q=0}$ we have $J_{\bf 0} = zJ$, which gives
the energy scale in the mean-field approximation.
Then the dimensionless integral $I$ gives the correction
to the mean-field critical temperature.
From integration over the appropriate Brilluoin zones, the values of
$I$ are $1.517$, $1.393$ and $1.343$ for sc, bcc and fcc lattices, respectively.
When applied to the 3D Ising model ($n=1$) one gets\cite{White83} 
$k_B T_c/JS^2 = z/I = 3.955, 5.743, 8.932 $ for sc, bcc and fcc
lattices, considerable improvements over the
standard mean-field results given from $k_B T_c/JS^2 = z$.
They compare favorably with the exact Ising model 
results\cite{Fisher67,AM76} from series: 
$k_B T_c/JS^2 = 4.5103, 6.3508, 9.794$, respectively.
For the Heisenberg model ($n=3$), the ORF predictions would
be $k_B T_c/JS^2 = 1.318, 1.914$, for sc and bcc lattices,
whereas precise Monte Carlo estimates\cite{Chen93} give
$k_B T_c/JS^2 = 1.443, 2.054$, respectively.
Note, however, that the ORF estimates are all below the exact 
results.

\section{Bravais Lattices with a Basis}
Now suppose the lattice has an underlying set of $N$ Bravais lattice
points ${\bf n}$, each of which has a two-atom basis $\{ {\bf 0}, {\bf d} \}$.
(The generalization to a larger basis is straightforward.)
Therefore at each site ${\bf n}$ we suppose there is a two component
field $w_{\bf n}$ written as a column vector:
\begin{equation}
w_{\bf n} = 
\left( \begin{array}{l} S_{\bf n} \\ S_{\bf n+d} \end{array} \right)
\end{equation}
It is useful to employ a sublattice notation, 
$S_{\bf n}\equiv S_{\bf n}^{A}, S_{\bf n+d}\equiv S_{\bf n}^{B}$.
The exchange interaction occurs between neighboring Bravais
sites, via a $2 \times 2$ matrix, $G_{\bf n,m}$:
\begin{equation}
{\cal H}_{\rm ex} = - \frac{1}{2}\sum_{\bf n}\sum_{\bf m} 
w_{\bf n}^{T}\cdot G_{\bf n,m} \cdot w_{\bf m},
\end{equation}
where it is stressed that the sums are over all Bravais sites; the
factor of $1/2$ avoids double counting, and
\begin{equation}
G_{\bf n,m} = 
\left( \begin{array}{ll}
G^{AA}  &  G^{AB} \\ G^{BA} & G^{BB} 
\end{array} \right)_{\bf n,m}
=
\left( \begin{array}{ll}
J_{\bf n,m}    &  J_{\bf n, m+d} \\
J_{\bf n+d, m} &  J_{\bf n+d, m+d} 
\end{array} \right)
\end{equation}
In fact, the matrix $G_{\bf n,m}$ is taken as zero unless ${\bf n-m}$
is a near neighbor displacement.
The details of the specific lattice will determine which components
of $G$ are nonzero.
%

The reaction terms, two for each site ${\bf n}$, can be written
with a $2\times 2$ unit matrix,
\begin{equation}
{\cal H}_{\rm orf} = \frac{1}{2} \sum_{\bf n} \sum_{\bf m} 
 w_{\bf n}^{T} \cdot \left[ 2 \lambda \delta_{n,m} 
\left( \begin{array}{cc} 1 & 0 \\ 0 & 1 \end{array} \right) \right] 
\cdot w_{\bf m}
\end{equation}
which is equivalent to expression (\ref{react}), and essentially shifts
the original exchange matrix by $-2\lambda \delta_{n,m} {\cal I}$, where 
${\cal I}$ is the $2\times 2$ unit matrix.
%

Finally, for the purpose of the calculation, there is a separate
applied field for each sublattice, so at a given site ${\bf n}$,
we have fields $H_{\bf n}^{A}$ (acting on $S_{\bf n}$) and 
$H_{\bf n}^{B}$ (acting on $S_{\bf n+d}$). 
These compose a column vector field,
\begin{equation}
h_{\bf n} = 
\left( \begin{array}{l} H_{\bf n}^{A} \\ H_{\bf n}^{B} \end{array} \right)
\end{equation}
%

Then we choose to write the q-space Hamiltonian, including the 
reaction term and applied fields, as
\begin{eqnarray}
{\cal H} = -\frac{1}{2}\sum_{\bf q} \bigl[  &&
w_{\bf -q}^{T}\cdot \bigl( G_{\bf q}-2\lambda {\cal I} \bigr)
\cdot w_{\bf q}  \nonumber \\
&& 
+ \bigl( w_{\bf -q}^{T}\cdot h_{\bf q} 
+ w_{\bf q}^{T}\cdot h_{\bf -q} \bigr) \bigr]
\end{eqnarray}
The Fourier transforms needed here obviously are related to
those already defined for $J_{\bf n-m}$, $S_{\bf n}$, etc.
The most significant difference from Eq.\ (\ref{Hamil-q}) is
the presence of ``2'' on $\lambda$, due to the two-atom basis.
This Hamiltonian is exact.
Now we define the sublattice magnetizations (where $i=A, B$),
\begin{equation}
M_{\bf q}^{i} = \langle S_{\bf q}^{i} \rangle
\end{equation}
and related susceptibilities (where also $j = A, B$),
\begin{equation}
\label{chi-ij}
\chi_{\bf q}^{ij} = \frac{\partial M_{\bf q}^{i}}{\partial H_{\bf q}^{j}}
=\beta \langle S_{\bf q}^{i} S_{\bf -q}^{j} \rangle,
\end{equation}
%

In the RPA, from the point of view of the negative q-components,
the Hamiltonian is approximated as
\begin{equation}
{\cal H} = - \sum_{\bf q} \left[ w_{\bf -q}^{T}\cdot h_{\bf q}
+ w_{\bf -q}^{T} \cdot K_{\bf q} \cdot \langle w_{\bf q} \rangle \right]
\end{equation}
where we use the shorthand notation for the shifted exchange interaction,
\begin{equation}
K_{\bf q} \equiv G_{\bf q} - 2\lambda {\cal I}
\end{equation}
The above Hamiltonian can alternatively be written in terms
of effective fields in components,
\begin{equation}
{\cal H} = - \sum_{\bf q} \left[ 
H_{\bf q}^{A ~ \rm eff} S_{\bf -q}^{A} +
H_{\bf q}^{B ~ \rm eff} S_{\bf -q}^{B} \right]
\end{equation}
\begin{mathletters}
\begin{equation}
H_{\bf q}^{A ~ \rm eff} = H_{\bf q}^{A} 
+ K_{\bf q}^{AA} \langle S_{\bf q}^{A} \rangle 
+ K_{\bf q}^{AB} \langle S_{\bf q}^{B} \rangle 
\end{equation}
\begin{equation}
H_{\bf q}^{B ~\rm eff} = H_{\bf q}^{B} 
+ K_{\bf q}^{BA} \langle S_{\bf q}^{A} \rangle 
+ K_{\bf q}^{BB} \langle S_{\bf q}^{B} \rangle
\end{equation}
\end{mathletters}
Then in the limit of zero applied field, using this RPA
Hamiltonian, the susceptibility definitions (\ref{chi-ij})
become
\begin{equation}
\chi_{\bf q}^{ij} = \frac{\partial M_{\bf q}^{i}}{\partial H_{\bf q}^{i \rm eff}}
\frac{\partial H_{\bf q}^{i \rm eff}}{\partial H_{\bf q}^{j}}
\end{equation}
and we get equations for the susceptibility components,
\begin{mathletters}
\label{chi-comp}
\begin{equation}
\chi_{\bf q}^{AA}=
\beta \langle S_{\bf q}^{A} S_{\bf -q}^{A} \rangle_{0}
\left[ 1 + K_{\bf q}^{AA} \chi_{\bf q}^{AA}
         + K_{\bf q}^{AB} \chi_{\bf q}^{BA} \right]
\end{equation}
\begin{equation}
\chi_{\bf q}^{AB}=
\beta \langle S_{\bf q}^{A} S_{\bf -q}^{A} \rangle_{0}
\left[ K_{\bf q}^{AA} \chi_{\bf q}^{AB}
     + K_{\bf q}^{AB} \chi_{\bf q}^{BB} \right]
\end{equation}
\begin{equation}
\chi_{\bf q}^{BB}=
\beta \langle S_{\bf q}^{B} S_{\bf -q}^{B} \rangle_{0}
\left[ 1 + K_{\bf q}^{BB} \chi_{\bf q}^{BB}
         + K_{\bf q}^{BA} \chi_{\bf q}^{AB} \right]
\end{equation}
\begin{equation}
\chi_{\bf q}^{BA}=
\beta \langle S_{\bf q}^{B} S_{\bf -q}^{B} \rangle_{0}
\left[ K_{\bf q}^{BB} \chi_{\bf q}^{BA}
     + K_{\bf q}^{BA} \chi_{\bf q}^{AA} \right]
\end{equation}
\end{mathletters}
where $\langle \quad \rangle_{0}$ means the expectation value using
the RPA Hamiltonian.
In the high-temperature limit, these expectations are
\begin{equation}
\beta \langle S_{\bf q}^{A} S_{\bf -q}^{A} \rangle_{0} = 
\beta \langle S_{\bf q}^{B} S_{\bf -q}^{B} \rangle_{0} = 
\frac{\beta}{n}S^2 \equiv \chi_0.
\end{equation}
The equations (\ref{chi-comp}) can be solved in the general case
for all four susceptibility components.
We get
\begin{mathletters}
\begin{equation}
\label{chi-AA}
\chi_{\bf q}^{AA} = \chi_{0} \frac{1-\chi_{0}K_{\bf q}^{BB}} 
{(1-\chi_{0}K_{\bf q}^{AA})(1-\chi_{0}K_{\bf q}^{BB})
 -\chi_{0}^{2}K_{\bf q}^{AB}K_{\bf q}^{BA}}
\end{equation}
\begin{equation}
\chi_{\bf q}^{BA} = \chi_{0} \frac{K_{\bf q}^{BA} \chi_{\bf q}^{AA}}
                                  {1-\chi_{0}K_{\bf q}^{BB}}
\end{equation}
\end{mathletters}
and similar equations for $\chi_{\bf q}^{BB}$ and $\chi_{\bf q}^{AB}$
by appropriately interchanging the AB indeces.
For the lattices considered in this paper, however, there
are the symmetries, $K_{\bf q}^{AA}=K_{\bf q}^{BB}$ and 
$K_{\bf q}^{AB}={(K_{\bf q}^{BA})}^{*}$.
Therefore the solutions are seen to satisfy symmetries 
$\chi_{\bf q}^{AA}=\chi_{\bf q}^{BB}$, 
$\chi_{\bf q}^{AB}={(\chi_{\bf q}^{BA})}^{*}$.
%

Now consider how to determine the reaction field $\lambda$,
and subsequently, $T_c$.
A little consideration using the definitions of $\chi_{\bf q}^{AA}$
and $S_{\bf q}^{A}$  shows that there is still the constraint,
\begin{equation}
\label{AA-constraint}
{1 \over N} \sum_{\bf q} \chi_{\bf q}^{AA} = {\beta \over n} S^{2} 
\equiv \chi_{0}
\end{equation}
Converting to the continuum limit, 
\begin{equation}
\label{continuum}
\left( \frac{V}{N} \right) \left( \frac{1}{2\pi} \right)^3
\int d^3 q ~ \chi_{\bf q}^{AA} = \chi_{0}
\end{equation}
This equation implicitly determines the reaction coupling $\lambda$
for any $T>T_c$.  
It is not clear how to get an explicit solution for $\lambda$ from it;
a solution for $\lambda(T)$ for a given lattice can be found numerically
(below).
%

Now, just as described for the Bravais lattice systems,
the critical temperature $T_c$ is the temperature at
which any of the susceptibility components, at ${\bf q=0}$,
diverges.
(The $\chi_{\bf q}^{ij}$ are well defined on the high
temperature side of $T_c$.)
So $T_c$ is determined as the point at which the denominator
of Eq.\ (\ref{chi-AA}) goes to zero, leading to a relation
between the critical temperature (via $\chi_{0}$) and the
critical reaction field, 
\begin{equation}
\label{lam-crit}
\chi_{0}^{-1}+2\lambda = G_{\bf 0}^{AA}+\sqrt{G_{\bf 0}^{AB}G_{\bf 0}^{BA}}
\end{equation}
Using this in the constraint equation (\ref{continuum}) together with
the result (\ref{chi-AA}) for $\chi_{\bf q}^{AA}$, gives the
general result when the symmetry $G_{\bf q}^{AA}=G_{\bf q}^{BB}$
holds,
\begin{mathletters}
\label{Tc-general}
\begin{equation}
k_B T_c = \frac{J_{0}S^2}{nI}
\end{equation}
\begin{equation}
I= 
\frac{V}{N} \int_{\rm BZ} \frac{d^3 q}{(2\pi)^3} ~
\frac{J_{0}( J_{0}-G_{\bf q}^{AA}) }
{(J_{0}-G_{\bf q}^{AA})^2 -G_{\bf q}^{AB}G_{\bf q}^{BA} }
\end{equation}
\end{mathletters}
where $J_{0}$ is the effective ${\bf q}={\bf 0}$ exchange strength,
\begin{equation}
\label{J-zero}
J_{0} \equiv G_{\bf 0}^{AA}+\sqrt{G_{\bf 0}^{AB}G_{\bf 0}^{BA}}
\end{equation}
The integral $I$ defined in this way again gives the correction 
to the mean-field prediction for $T_c$.
Thus the determination of $T_c$ has been reduced to evaluating this
integral over the Brilluoin zone.

Below we will use Eq.\ (\ref{Tc-general}) to
estimate the critical temperature for diamond and hcp lattices.
But first we can verify that the result is correct by using it for 
a bcc lattice, considered as a simple-cubic lattice with a basis,
where we already know the standard ORF result for $T_c$.

\section{bcc Lattice as sc with Basis}
\label{bccLattice}
The bcc lattice points can be generated from the simple
cubic primitive vectors, ${\bf a}_1=a\hat{x}$, ${\bf a}_2=a\hat{y}$, 
and ${\bf a}_3=a\hat{z}$, where $a$ is the cubic cell lattice constant,
together with the basis, $\{{\bf 0}, {\bf d} \}$, 
where ${\bf d}=\frac{a}{2}(\hat{x}+\hat{y}+\hat{z})$ is
the displacement to the body-centered point.
The lattice can be thought of as a pair of interpenetrating 
sc lattices (A, B sublattices) with displacement ${\bf d}$.
The near neighbors of an `A' site are all `'B' sites,
and vice-versa, with the result that the $G^{AA}$ and
$G^{BB}$ couplings are all zero.
The nonzero $G_{\bf n,m}^{ij} \equiv G_{\bf r}^{ij}$ couplings 
depend only on the near neighbor displacement, ${\bf r}\equiv {\bf m-n}$, 
as follows:
\begin{mathletters}
\begin{eqnarray}
G_{\bf r}^{BA}=J, ~
{\bf r}  & = & {\bf 0},~ {\bf a}_1,~ {\bf a}_2,~ {\bf a}_3,~
{\bf a}_1+{\bf a}_2+{\bf a}_3, \nonumber \\
&& {\bf a}_1+{\bf a}_2,~ {\bf a}_2+{\bf a}_3,~ {\bf a}_3+{\bf a}_1
\end{eqnarray}
\begin{eqnarray}
G_{\bf r}^{AB}=J, ~
{\bf r}  & = & {\bf 0},~ -{\bf a}_1,~ -{\bf a}_2,~ -{\bf a}_3,~
-{\bf a}_1-{\bf a}_2-{\bf a}_3, \nonumber \\
&& -{\bf a}_1-{\bf a}_2,~ -{\bf a}_2-{\bf a}_3,~ -{\bf a}_3-{\bf a}_1
\end{eqnarray}
\end{mathletters}
where the terms for ${\bf r}={\bf 0}$ correspond to the coupling within 
the two-atom basis.
The fact that $G_{\bf n,m}^{AA}=G_{\bf n,m}^{BB}=0$ 
simplifies the determination of $T_c$ considerably, as
we only need to know the product, $G_{\bf q}^{AB}G_{\bf q}^{BA}$.
%

The Fourier-transformed interactions are found from sums over all
the nonzero $G_{\bf r}$ (15 possible terms):
\begin{equation}
\label{Gq}
G_{\bf q}^{AB} = \sum_{\bf r} G_{\bf r}^{AB} e^{i {\bf q}\cdot {\bf r} }
={(G_{\bf q}^{BA})}^{*}
\end{equation}
\begin{eqnarray}
G_{\bf q}^{AB} = J  & \{ & 1 + e^{-iq_x} + e^{-iq_y} + e^{-iq_z} \nonumber \\
&& + e^{-i(q_x+q_y)} + e^{-i(q_y+q_z)} + e^{-i(q_z+q_x)} \nonumber \\
&& + e^{-i(q_x+q_y+q_z)} \}
\end{eqnarray}
In this and the following equations, $q_x, q_y, q_z$ are in units of
$1/a$.
For the underlying sc lattice, the density of points is 1 for every volume
$a^3$, i.e., $V/N=a^3$.
After a short calculation, there results
\begin{eqnarray}
G_{\bf q}^{AB}&&G_{\bf q}^{BA} = 
8J^2 \left\{ 1 + \cos q_x  + \cos q_y  + \cos q_z  \right. \nonumber \\
&& + \cos q_x ~ \cos q_y + \cos q_y ~ \cos q_z + \cos q_z ~ \cos q_x \nonumber \\ 
&& \left. + \cos q_x ~ \cos q_y ~ \cos q_z \right\}
\end{eqnarray} 
and the determination of $T_c$ relies on evaluation of the simplified 
integral,
\begin{equation}
\label{I-simple}
I= 
\int \frac{d^3q}{(2\pi)^3} ~ 
\frac{J_{0}^2}{J_{0}^2-G_{\bf q}^{AB}G_{\bf q}^{BA} }
\end{equation}
where $J_{0}=G_{\bf 0}^{AB}=G_{\bf 0}^{BA}=8J=zJ$ and the
factor $\frac{V}{N}=a^3$ was absorbed into the dimensionless $q$.
\begin{eqnarray}
\label{I-sc}
I &=& 
\int_{-\pi}^{\pi} \frac{dq_x}{2\pi} ~ \int_{-\pi}^{\pi} \frac{dq_y}{2\pi} ~ 
\int_{-\pi}^{\pi} \frac{dq_z}{2\pi} ~ \nonumber \\
&& \Bigl\{ 
1-{1 \over 8} \bigl[ 1 + \cos q_x  + \cos q_y  + \cos q_z  \nonumber \\
&& + \cos q_x ~ \cos q_y + \cos q_y ~ \cos q_z + \cos q_z ~ \cos q_x \nonumber \\ 
&& + \cos q_x ~ \cos q_y ~ \cos q_z \bigr] \Bigr\}^{-1}
\end{eqnarray}
The integral $I$ was evaluated numerically by sampling $q_x, q_y, q_z$
uniformly with a constant increment, and then using an extrapolation
of the results in the limit that the increment goes to zero.
Essentially, we let $dq_x=dq_y=dq_z=2\pi/N_x$, where $N_x$ is some
integer number of divisions of the axes, and then generated 
a simple cubic lattice of sampling points from these.
The integral $I$ was then estimated as a sum over the resulting
cubic grid of sampling points in the BZ in q-space.
A plot of $I$ versus $1/N_x$ results in a straight line whose
extrapolation to $1/N_x \rightarrow 0$ gives a very accurate estimate
of the integral. (Errors in the integral estimate clearly go
as $1/N_x$.)
In this way we found $I=1.39321$.
Therefore, the estimate of critical temperature for the bcc system 
obtained viewing it as sc with a basis is
\begin{equation}
\label{Tc-bcc}
k_B T_c = \frac{8JS^2}{n} \frac{1}{1.39321} = \frac{5.74213}{n} JS^2.
\end{equation}
%

For comparison with the standard ORF procedure, when the original bcc
lattice is used, the specific volume is $V/N=\frac{1}{2}a^3$.
The Fourier transformed exchange (Eq.\ \ref{Jq}) is
\begin{equation}
J_{\bf q}=8J \cos\frac{q_x}{2} ~ \cos\frac{q_y}{2} ~ \cos\frac{q_z}{2}
\end{equation}
The integral needed for Eq.\ (\ref{Tc-general}) is
\begin{eqnarray}
\label{I-bcc}
I &=& \frac{1}{2} \int_{\rm BZ} \frac{d^3 q}{(2\pi)^3} 
\Bigl\{ 
1 - \cos\frac{q_x}{2} ~ \cos\frac{q_y}{2} ~ \cos\frac{q_z}{2} 
\Bigr\}^{-1} \nonumber \\
&=& \frac{1}{2} \times 2.786 = 1.393,
\end{eqnarray}
where the integral is over the Brilluoin zone for the bcc lattice
(an fcc Wigner-Seitz cell), and was evaluated by the sampling
technique already described.
The integral is exactly twice that given by expression (\ref{I-sc}),
but because $V/N=\frac{1}{2}a^3$ is half as large for the bcc lattice as 
for the sc lattice, the result for $T_c$ from Eq.\ (\ref{Tc-general})
is that given by Eq.\ (\ref{Tc-bcc}).
Thus this approach using the two-atom basis, for the bcc system, is 
equivalent to the standard ORF procedure, and should be reliable
for application to the diamond and hcp lattices.

As an interesting mathematical note, the integral $I$ in (\ref{I-bcc})
can be rewritten using the periodicity and symmetry in q-space to be
over the same region as in (\ref{I-sc}).
A cubic cell $-2\pi\le q_x \le 2\pi, -2\pi\le q_z \le 2\pi,
-2\pi\le q_z \le 2\pi$, contains 4 copies of the Brilluoin zone
of the bcc lattice.
Thus we can integrate over this cubic cell and divide by 4;
also, shifting the variables of integration to $q^{\prime}_x = q_x/2$,
etc., leads to an additional factor of $2^3$ and gives 
\begin{eqnarray}
\label{I-bcc-prime}
I = \frac{1}{2} \times \frac{1}{4} \times 2^3 &&
\int_{-\pi}^{\pi} \frac{dq^{\prime}_x}{2\pi} ~ 
\int_{-\pi}^{\pi} \frac{dq^{\prime}_y}{2\pi} ~ 
\int_{-\pi}^{\pi} \frac{dq^{\prime}_z}{2\pi} ~ 
\nonumber \\
&&
\Bigl\{ 
1 - \cos q^{\prime}_x ~ \cos q^{\prime}_y ~ \cos q^{\prime}_z 
\Bigr\}^{-1} 
\end{eqnarray}
Overall, the prefactor on the integral is one.
Thus the integral here must be the same 
as the integral in (\ref{I-sc}).
The integrands, however, are not equivalent;  there is no way
to transform one into the other.
%

Finally we also note that it is easier and more precise to evaluate
(\ref{I-bcc-prime}) by a uniform cubic grid of sampling points than  
expression (\ref{I-bcc}), because a cubic grid of sampling points
easily fits to the cubic integration boundaries.
On the other hand, for expression (\ref{I-bcc}), there is always
more difficulty to fit any grid of points smoothly to {\em all} of the
integration boundaries of the fcc Wigner-Seitz cell in q-space,
leading to greater discretization errors.
Removing these errors is important for producing a smooth extrapolation 
to $N_x\rightarrow\infty$.

\section{Diamond Lattice}
\begin{figure}
  \hskip1.6in 
  \psfig{figure=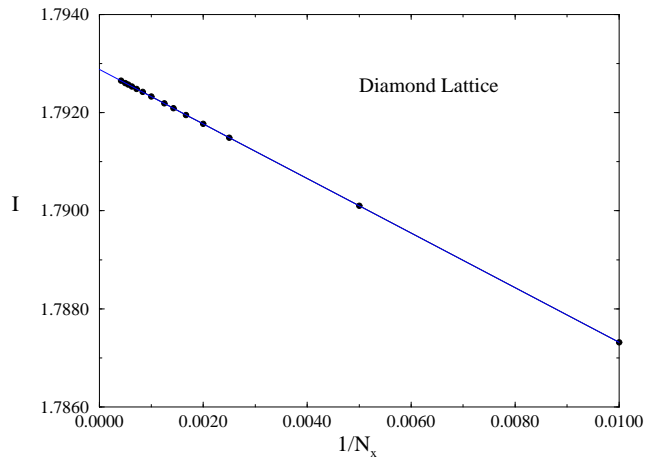,angle=-90.0,width=3.5in}
\caption{
\label{I-diamond}
Evaluation of the integral $I$ of Eq.\ \protect\ref{I.diamond} for
the diamond lattice, for different numbers of grid points $N_x$ along 
each axis.  The errors go as $1/N_x$ and extrapolation to 
$N_x\rightarrow\infty$ gives $I=1.79288$.
}
\end{figure}

The diamond lattice can be considered as an fcc lattice
with a two atom basis. 
The fcc primitive vectors are 
${\bf a}_1=\frac{a}{2}(\hat{x}+\hat{y}),
{\bf a}_2=\frac{a}{2}(\hat{y}+\hat{z}), 
{\bf a}_2=\frac{a}{2}(\hat{z}+\hat{x})$, 
where $a$ is the standard cubic cell lattice constant,
and the basis is $\{{\bf 0}, {\bf d} \}$, 
where ${\bf d}=\frac{a}{4}(\hat{x}+\hat{y}+\hat{z})$.
The nearest neighbors of a site ${\bf n+d}$ (a `B'-site)
are $\{ {\bf n}, {\bf n}+{\bf a}_1, {\bf n}+{\bf a}_2, {\bf n}+{\bf a}_3 \}$
(all `A'-sites).
The nearest neighbors of a site ${\bf n}$ (an `A'-site)
are at $\{ {\bf n}+{\bf d}, {\bf n}+{\bf d}-{\bf a}_1, 
{\bf n}+{\bf d}-{\bf a}_2, {\bf n}+{\bf d}-{\bf a}_3 \}$
(all `B'-sites).
As a result, only $G^{AB}$ and $G^{BA}$ are nonzero.
In terms of the neighbor displacements ${\bf r}={\bf m-n}$,
the only nonzero exchange couplings are
\begin{mathletters}
\begin{equation}
G_{\bf r}^{BA}=J, ~ 
{\bf r}=\{ {\bf 0}, {\bf a}_1, {\bf a}_2, {\bf a}_3 \}
\end{equation}
\begin{equation}
G_{\bf r}^{AB}=J, ~
{\bf r}=\{ {\bf 0}, -{\bf a}_1, -{\bf a}_2, -{\bf a}_3 \}
\end{equation}
\end{mathletters}
Then it is straightforward to evaluate 
\begin{mathletters}
\begin{eqnarray}
G_{\bf q}^{AB} &=& \sum_{\bf r} G_{\bf r}^{AB} e^{i {\bf q} \cdot {\bf r} } 
= (G_{\bf q}^{BA})^{*} \\
&=& J \left[ 1+e^{ -i{\bf q} \cdot {\bf a}_1 }
+e^{ -i{\bf q} \cdot {\bf a}_2 }
+e^{ -i{\bf q} \cdot {\bf a}_3 } \right]
\end{eqnarray}
\end{mathletters}
and what is needed for the $T_c$ integral:
\begin{eqnarray}
G_{\bf q}^{AB}G_{\bf q}^{BA} = 
4J^2 & \Bigl[ & 1 + \cos\frac{q_x}{2} \cos\frac{q_y}{2} 
\nonumber \\
&& 
+ \cos\frac{q_y}{2} \cos\frac{q_z}{2}
+ \cos\frac{q_z}{2} \cos\frac{q_x}{2}  \Bigr]
\end{eqnarray}
where $q$'s are in units of $1/a$.
Clearly we also have once again, 
$J_{0}=G_{\bf 0}^{AB}=G_{\bf 0}^{BA}=4J = zJ$.
The specific volume per lattice point on the underlying fcc lattice 
is $V/N=\frac{1}{4}a^3$.
Then $T_c$ will be evaluated using Eq.\ (\ref{Tc-general}) and
the simplified integral like (\ref{I-simple}), which becomes
\begin{eqnarray}
I  =  \frac{1}{4} && \int_{\rm BZ} \frac{d^3 q}{(2\pi)^3} \nonumber \\
&&
\Bigl\{ 
1-\frac{1}{4}\Bigl[ 1 + \cos\frac{q_x}{2} \cos\frac{q_y}{2} 
\nonumber \\
&&
+ \cos\frac{q_y}{2} \cos\frac{q_z}{2}
+ \cos\frac{q_z}{2} \cos\frac{q_x}{2}  \Bigr] 
\Bigr\}^{-1}
\end{eqnarray}
The BZ for the fcc lattice is a bcc Wigner-Seitz cell, with lattice
constant $4\pi/a$.
The easiest and most precise way to evaluate this integral is
to apply the same procedure we noted for the bcc system:
Change the integration region to the cube,
$-2\pi \le q_x \le 2\pi, -2\pi \le q_y \le 2\pi, -2\pi \le q_z \le 2\pi$,
which contains 2 copies of the BZ, therefore include a factor of $\frac{1}{2}$,
and sum over points on a cubic grid.
Also make the variable change, ${\bf q}^{\prime} = {\bf q}/2$,
which leads to a factor of $2^3$, giving
\begin{eqnarray}
\label{I.diamond}
I  =  \frac{1}{4} \times && \frac{1}{2} \times  2^3 
\int_{-\pi}^{\pi} \frac{dq^{\prime}_x}{2\pi} ~ 
\int_{-\pi}^{\pi} \frac{dq^{\prime}_y}{2\pi} ~ 
\int_{-\pi}^{\pi} \frac{dq^{\prime}_z}{2\pi} ~ 
\nonumber \\
&&
\Bigl\{ 
1-\frac{1}{4}\Bigl[ 1 + \cos q^{\prime}_x \cos q^{\prime}_y 
\nonumber \\
&&
+ \cos q^{\prime}_y \cos q^{\prime}_z 
+ \cos q^{\prime}_z \cos q^{\prime}_x  \Bigr] \Bigr\}^{-1}
\end{eqnarray}
Using a cubic grid spacing $dq_x^{\prime}=dq_y^{\prime}=dq_z^{\prime}=2\pi/N_x$,
we evaluated the integral as a sum over points within the cubical
integration region, for $N_x$ ranging from 100 to 2000.  
A plot of $I$ versus $1/N_x$ gives a straight line (Fig.\ \ref{I-diamond}),  
and its extrapolation to $1/N_x \rightarrow 0$ gives $I=1.79288 $.
Thus the critical temperature is estimated as
\begin{equation}
\label{Tc-diamond}
k_B T_c = \frac{4JS^2}{n} \frac{1}{1.79288} = \frac{2.23105}{n} JS^2
\end{equation} 
The exact result for $T_c$ (Ising model, $n=1$) as estimated from series 
expansions, is known to be $k_B T_c = 2.7040 JS^2$.
Thus the ORF calculation, as is usual, underestimates $T_c$ but
is a considerable improvement over the simple mean-field
result, $k_B T_c = 4 JS^2$.

\section{Simple Hexagonal Bravais Lattices and HCP Lattices}
Another example of a lattice with a basis is the hcp system,
which can be considered as interpenetrating simple hexagonal
Bravais lattices\cite{AM76} (i.e., stacked triangular nets).
The primitive vectors of the simple hexagonal Bravais lattice
can be taken as ${\bf a}_1=a\hat{x}, 
{\bf a}_2=a(\frac{1}{2}\hat{x}+\frac{\sqrt{3}}{2}\hat{y}),
{\bf a}_3=c\hat{z}$, where $a$ and $c$ are the lattice constants.
For the hcp system, a two-atom basis of 
$\{ {\bf 0}, {\bf d} \}$ is used, where
${\bf d}=\frac{1}{3}({\bf a}_1+{\bf a}_2)+\frac{1}{2}{\bf a}_3$, and
one triangular net is stacked on top of the previous
one, but shifted to be over the centers of one set of the triangular 
cells below, in what is usually referred to as the ABAB... packing.
For the lattice constant ratio $c/a=\sqrt{8/3}$, the highest 
density packing is obtained, however, for the calculation here this
ratio does not directly enter, and need not be specified.
Instead, it is interesting to consider that the near neighbor
exchange interactions within the triangular nets (xy-plane) 
have one strength, $J_{xy}$, while there is a different strength,
$J_{z}$, for the bonds between the planes.
In general we can consider the calculation of $T_c$ as a function
of the ratio, $\Delta \equiv J_{z}/J_{xy}$.
%

We present first the calculation of $T_c(\Delta)$ for the simple 
hexagonal Bravais lattice, using the standard ORF theory in 
Sec.\ \ref{Bravais}, which acts as an introduction to the 
corresponding calculation for the hcp system, because they both 
rely on the same information concerning the Brilluoin zone. 

\subsection{Simple Hexagonal Bravais Lattice}
\label{SHBL}
Here there are 6 neighbor displacements from some arbitrary site to
neighbors in the same plane 
$\{ \pm {\bf a}_1, \pm {\bf a}_2, \pm ({\bf a}_1-{\bf a}_2) \}$,
with exchange strength, $J_{xy}$.
The remaining two neighbors, with displacements, 
$ \pm {\bf a}_3$, have exchange strengths, $J_{z}$.
A short calculation shows that the q-space exchange (Eq.\ \ref{Jq}) is
\begin{mathletters}
\begin{eqnarray}
J_{\bf q} &=& 2J_{xy} \bigl[ 
\cos {\bf q}\cdot {\bf a}_1 + \cos {\bf q}\cdot {\bf a}_2
+ \cos {\bf q}\cdot ({\bf a}_1-{\bf a}_2)  \bigr]
\nonumber \\
&& +2J_{z} \cos {\bf q}\cdot {\bf a}_3 \\
&=& 2J_{xy} \bigl[ 
\cos q_x a + 2 \cos \frac{1}{2}q_x a \cos \frac{\sqrt{3}}{2}q_y a \bigr]
\nonumber \\
&&
+2J_{z} \cos q_z c
\end{eqnarray}
\end{mathletters}
and $J_{\bf 0}=6J_{xy}+2J_{z}=2(3+\Delta)J_{xy}$ will determine
the mean-field critical temperature.
%

\begin{figure}
  \hskip1.6in 
  \psfig{figure=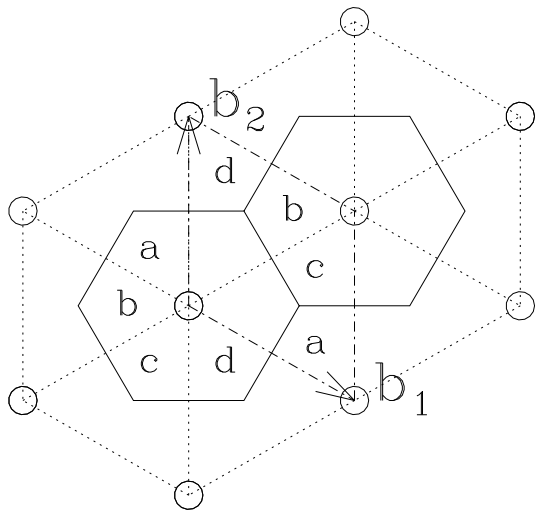,angle=0.0,width=3.5in}
\caption{
\label{BZ-hex}
Wigner-Seitz cells (solid line hexagons) for the simple hexagonal Bravais lattice 
reciprocal space, compared with the equivalent rhombic cell (dot-dash) used 
for integrals.  Segments labeled {\it a,b,c,d} are equivalent by symmetry 
operations. 
}
\end{figure}

The area of one triangle in the net is 
$\frac{1}{2}a\times \frac{\sqrt{3}}{2}a=\frac{\sqrt{3}}{4}a^2$,
and there is $\frac{1}{2}$-site per triangle per layer.
Thus the specific volume per site is $V/N=\frac{\sqrt{3}}{2}a^2 c$.
The primitive vectors of the reciprocal space are
\begin{mathletters}
\label{hex-b-vecs}
\begin{equation}
{\bf b}_1=\frac{4\pi}{\sqrt{3}a} 
~ \bigl( \frac{\sqrt{3}}{2}\hat{x} - \frac{1}{2}\hat{y} \bigr)
\end{equation}
\begin{equation}
{\bf b}_2=\frac{4\pi}{\sqrt{3}a} ~ \hat{y}
\end{equation}
\begin{equation}
{\bf b}_3=\frac{2\pi}{c} ~ \hat{z}
\end{equation}
\end{mathletters}
The reciprocal space is another simple hexagonal lattice,
with lattice constants $\frac{4\pi}{\sqrt{3}a}$ in the
xy-plane and $\frac{2\pi}{c}$ in the z-direction,
rotated by $30^{o}$ from the real space lattice.
The Brilluoin zone Wigner-Seitz cell is a hexagonal
cylinder, however, for the purpose of the integral needed
here (Eq.\ \ref{ORF-I}) it is more convenient to do 
the summation inside
a cell bounded in the xy-plane by a rhombus formed by
${\bf b}_1$ and ${\bf b}_2$ (See Fig. \ref{BZ-hex}.).
The hexagonal cylinder and rhombical cylinder cells have equal areas
and are equivalent to each other by appropriate symmetry operations.
This rhombical cylinder cell is very convenient for evaluation of
the integral $I$, especially with the variable change on ${\bf q}$:
\begin{equation}
\label{var-change}
{\bf q}=x{\bf b}_1+y{\bf b}_2+z{\bf b}_3
\end{equation} 
where the dimensionless parameters $x,y,z$ all range from $0$ to $1$,
mapping out the entire cell.
This leads to 
\begin{equation}
\label{vol-element}
d^3 q = {\bf b}_1 \cdot ( {\bf b}_2 \times {\bf b}_3 ) ~dx ~ dy ~ dz  
= (2\pi)^3 \frac{N}{V}  ~dx ~ dy ~ dz
\end{equation}
and the integral is simplified to,
\begin{eqnarray}
I &=& \int_{0}^{1} dx \int_{0}^{1} dy \int_{0}^{1} dz ~
\Bigl\{ 1-\frac{1}{3+\Delta}\Bigl[ \cos 2\pi x 
\nonumber \\
&&
+ \cos 2\pi y + \cos 2\pi (x-y) + \Delta \cos 2\pi z \Bigr] 
\Bigr\}^{-1}
\end{eqnarray}
The integral gives the correction to the mean-field prediction, i.e.,
\begin{equation}
k_B T_c = \frac{k_B T_c^{MF}}{I}, \quad k_B T_c^{MF} = \frac{2(3+\Delta)J_{xy}S^2}{n}.
\end{equation}
%

\begin{figure}
  \hskip1.6in 
  \psfig{figure=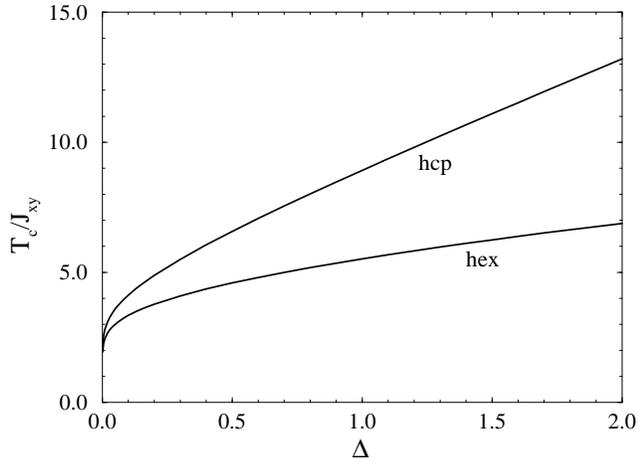,angle=-90.0,width=3.5in}
\caption{
\label{Fig-Tc-hcphex}
ORF results for $T_c$ on the simple hexagonal Bravais lattice (hex)
and the hexagonal close packed lattice (hcp), as functions of 
the exchange anisotropy $\Delta=J_{z}/J_{xy}$.
}
\end{figure}

The integral $I$ was evaluated by the numerical techniques
described above (Sec.\ \ref{bccLattice}), for a range of anisotropy
parameter $0 < \Delta \le 2$, see Fig.\ \ref{Fig-Tc-hcphex}.
At the isotropic limit, $\Delta=1$, we get $I=1.44930$ and $n k_B T_c =5.5199 JS^2$.
In the limit $\Delta\rightarrow 0$, the system becomes two-dimensional,
the integral $I$ diverges logarithmically due to small-${\bf q}$
contributions, and ORF is not applicable.
%

\subsection{Hexagonal Close Packed Lattices}
\begin{figure}
  \hskip1.6in
  \psfig{figure=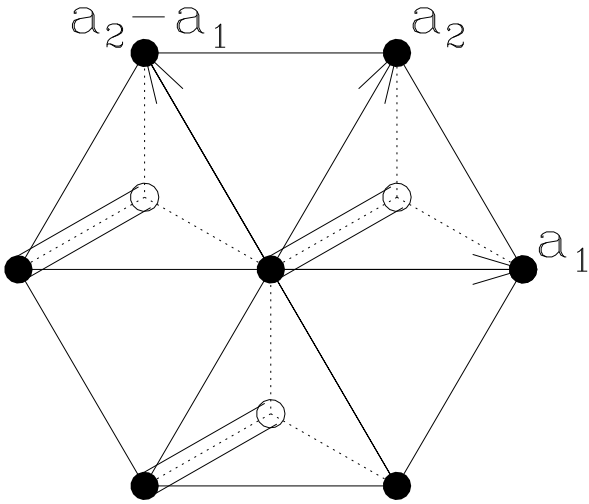,angle=0.0,width=3.5in}
\caption{
\label{hcp-diagram}
XY projection of some nearest-neighbor bonds between A-sites (solid circles)
and B-sites (open circles) in an hcp lattice.  Double solid lines connect 
A, B sites at the same Bravais lattice point.  Solid lines show AA bonds 
(within the planes), dotted lines show AB bonds.
}
\end{figure}

Again there are 6 neighbor displacements from an arbitrary site to
neighbors in the same plane 
$\{ \pm {\bf a}_1, \pm {\bf a}_2, \pm ({\bf a}_1-{\bf a}_2) \}$,
with exchange strength, $J_{xy}$.
The difference from the simple hexagonal lattice, is that there
are 3 neighbors in a layer above and 3 neighbors in a layer
below the one being considered, with exchange couplings $J_{z}$,
giving 12 neighbors in all.
However, to evaluate the matrix elements $G^{i,j}_{\bf r}$, 
we need to consider these couplings from the point of view
of the simple hexagonal Bravais lattice with a basis
(see Fig.\ \ref{hcp-diagram}).
Thus, an arbitrary A-site, has the 6 neighbor displacements
to other A-sites in the same xy-plane: 
$\{ \pm {\bf a}_1, \pm {\bf a}_2, \pm ({\bf a}_1-{\bf a}_2) \}$.
which will give nonzero $G^{AA}$ coupling terms.
The neighbors in adjacent planes are B-sites, leading to
nonzero $G^{AB}$ terms.
Further consideration leads to the nonzero coupling elements,
\begin{mathletters}
\begin{equation}
G_{\bf r}^{AA}=G_{\bf r}^{BB}=J_{xy}, ~ 
{\bf r}=\pm {\bf a}_1, \pm {\bf a}_2, \pm ({\bf a}_1-{\bf a}_2 ) 
\end{equation}
\begin{equation}
G_{\bf -r}^{AB}=G_{\bf r}^{BA}=J_{z}, ~
{\bf r}={\bf 0}, {\bf a}_1, {\bf a}_2, 
{\bf a}_3, {\bf a}_3+{\bf a}_1, {\bf a}_3+{\bf a}_2
\end{equation}
\end{mathletters}
It is notable that it is the first example where the diagonal
elements are nonzero.
The terms where ${\bf r}={\bf 0}$ are the coupling within the basis.
%

The q-space couplings (Eq.\ \ref{Gq}) are found to be:
\begin{eqnarray}
G_{\bf q}^{AA}=G_{\bf q}^{BB}=2J_{xy} \bigl[ &&
\cos {\bf q}\cdot {\bf a}_1 + \cos {\bf q}\cdot {\bf a}_2
\nonumber \\
&&
+ \cos {\bf q}\cdot ({\bf a}_1-{\bf a}_2)  \bigr]
\end{eqnarray}
\begin{eqnarray}
G_{\bf q}^{AB}=(G_{\bf q}^{BA})^{*}= 
J_{z} \bigl[ && 1+e^{-i{\bf q}\cdot {\bf a}_1}
+e^{-i{\bf q}\cdot {\bf a}_2}
+e^{-i{\bf q}\cdot {\bf a}_3} 
\nonumber \\
&&
+e^{-i{\bf q}\cdot ({\bf a}_3+{\bf a}_1)}
+e^{-i{\bf q}\cdot({\bf a}_3+{\bf a}_2)} \bigr]
\end{eqnarray}
and what is needed for evaluation of $T_c$:
\begin{eqnarray}
G_{\bf q}^{AB}G_{\bf q}^{BA} &=& J_{z}^2 
\bigl[ 6+4 \{ \cos {\bf q}\cdot {\bf a}_1 + \cos {\bf q}\cdot {\bf a}_2 
\nonumber \\
&&
+\cos {\bf q}\cdot ({\bf a}_1-{\bf a}_2 ) \} \bigr]
\times \bigl[ 1+\cos {\bf q}\cdot {\bf a}_3 \bigr]
\end{eqnarray}
The q=0 exchange strength (Eq.\ \ref{J-zero}) is seen to be 
$J_{\bf 0}=6(J_{xy}+J_{z})=6 J_{xy}(1+\Delta)$ and determines
the $\Delta$-dependent mean-field critical temperature.
The specific volume per site (for the underlying simple hexagonal
Bravais lattice) is $V/N=\frac{\sqrt{3}}{2}a^2 c$.
The reciprocal space is that of the simple hexagonal Bravais lattice
as described in Sec.\ \ref{SHBL}.
Therefore, using the rhombical cylinder Brilluoin zone cell, 
the integral $I$ of Eq.\ \ref{Tc-general} can be transformed 
using Eqs.\ \ref{var-change} and \ref{vol-element} into
\begin{equation}
\label{I_hcp}
I = \int_{0}^{1} dx \int_{0}^{1} dy \int_{0}^{1} dz ~
\frac{J_{0}( J_{0}-G_{\bf q}^{AA}) }
{(J_{0}-G_{\bf q}^{AA})^2 -G_{\bf q}^{AB}G_{\bf q}^{BA} }
\end{equation}
where we have the transformed quantities
\begin{mathletters}
\begin{equation}
G_{\bf q}^{AA} = 2J_{xy} \Bigl[ \cos 2\pi x + \cos 2\pi y + \cos 2\pi (x-y) \Bigr]
\end{equation}
\begin{eqnarray}
G_{\bf q}^{AB} && G_{\bf q}^{BA} = 2 J_{xy} \Delta 
\Bigl[ 1 + \cos 2\pi z \Bigr] \times 
\nonumber \\
&&
\Bigl[ 3 + 2 \Bigl\{ \cos 2\pi x + \cos 2\pi y + \cos 2\pi (x-y) \Bigr\} \Bigr]
\end{eqnarray}
\end{mathletters}
The correction to the mean-field prediction is
\begin{equation}
k_B T_c = \frac{k_B T_c^{MF}}{I}, \quad k_B T_c^{MF} = \frac{6(1+\Delta)J_{xy}S^2}{n}.
\end{equation}
%

$I$ was evaluated by the numerical techniques
described above (Sec.\ \ref{SHBL}), including the $N_x\rightarrow\infty$
extrapolation. 
For example, at the isotropic limit, $\Delta=1$, we get $I=1.34466$, and 
$n k_B T_c=8.92418 JS^2$.
It is interesting to note that the same value of $I$ results for
the fcc lattice (Eq.\ \ref{ORF-I}), when evaluated to the same precision.
Thus the ORF corrections to the mean-field $T_c$ for fcc and hcp
lattices, both with 12 nearest neighbors, are the same.
Some other hcp results in a limited range of anisotropy $\Delta$ are shown 
in Fig.\ \ref{Fig-Tc-hcphex}.
Once again, in the limit $\Delta\rightarrow 0$ there is a weak
divergence of the integral as the system crosses over into a
two-dimensional one, with $T_c\rightarrow 0$ over a very narrow
range of $\Delta$.
ORF is not applicable in this limit; $T_c$ should pass over
to the finite value for the 2D triangular lattice model.\cite{Wu82} 

There are a few theoretical results to compare with for the hcp system.
In a series of papers, Domany, Gubernatis and 
Auerbach\cite{Domany84,Domany+85,Auerbach+88}
analyzed a Lifshitz tricritical point for the hcp Ising model,
which occurs at a negative value of $\Delta$.
As part of their analysis they applied Monte Carlo 
calculations\cite{Domany+85} to determine the phase diagram;
very roughly for $\Delta=1$ they obtained $k_B T_c\approx 10 JS^2$.
Values of $T_c$ at other anisotropies also can be estimated from 
their Fig.\ 1 but with poor precision.
However, it does appear that the Onsager results fall below
the Monte Carlo estimates of $T_c$, as expected.

For the hcp Heisenberg model, Adler\cite{Adler81}
estimated $T_c$ by a Green's function approach
together with a random phase approximation. 
It is surprising to see that at $\Delta=1$ Alder found
the correction to the mean-field $T_c$ to be by the
factor $F(-1)=1.34\pm 0.005$, where $F(-1)$ is a
certain sum over the Brillouin zone.
Making a more accurate evaluation of $F(-1)$ by the techniques
described here, we get $F(-1)=1.34466$, i.e.,
a value exactly equal to the correction integral $I$ obtained
from the ORF procedure.\cite{AdlerNote}  
In fact, it is easy to show that the expression for $F(-1)$ given by
Adler is exactly equivalent to our expression (\ref{I_hcp}) for $I$,
including the anisotropic case, $\Delta\ne 1$.
Therefore, the Green's function approach used there is exactly
equivalent to the ORF procedure presented here; they
are different approaches to impose the random phase approximation.

Furthermore,  in this level of approximation, the
question posed by Domb and Sykes\cite{Domb+57} and investigated by
Adler is answered:  $T_c$ for fcc and hcp Ising models 
are the same, even though the hcp lattice is more densely packed
and might be expected to have a higher $T_c$.
Apparently a more precise procedure is needed to determine
whether there is a true difference in their critical temperatures.

\section{Reaction Field and Thermodynamic Quantities at $T>T_c$}
It is clear that any quantities such as specific heat, magnetization,
etc, can be evaluated via the RPA Hamiltonian for temperatures away 
from $T_c$, provided that the reaction field, $\lambda$, has been determined.
Thus we take a few sentences to examine how $\lambda$ can be calculated.
%

At the critical temperature, the reaction field as determined
from Eq.\ (\ref{lam-crit}) is seen to be
\begin{equation}
\lambda_c \equiv \lambda(T_c) =  \frac{1}{2}(J_{\bf 0}-\chi_{0}^{-1})
=\frac{1}{2}(J_{\bf 0}-n k_{B}T_c/S^2)
\end{equation}
For higher temperatures, the constraining equation (\ref{continuum}) to 
determine $\lambda$ is equivalent to
\begin{equation}
\frac{V}{N} \int_{BZ} \frac{d^3 q}{(2\pi)^3} ~ 
\frac{1-\chi_0 (G_{\bf q}^{AA}-2\lambda)}
{\bigl[ 1-\chi_0 (G_{\bf q}^{AA}-2\lambda) \bigr]^2 
         - \chi_{0}^2 G_{\bf q}^{AB}G_{\bf q}^{BA} } = 1
\end{equation}
Considering the left hand side as a function of $\lambda$,
one can apply Newton's method to search for the $\lambda$ at which 
the function passes through 1.  
This search can be aided by the requirement that the denominator
of this integrand must be positive everywhere in the BZ, including 
at ${\bf q}=0$.
This leads to the inequality for $T>T_c$,
\begin{equation}
\lambda(T)>\frac{1}{2}(J_{\bf 0}-\chi_{0}^{-1})
=\frac{1}{2}(J_{\bf 0}-n k_{B} T/S^2) < \lambda_c
\end{equation}
Indeed, we have found as well that $\lambda(T)<\lambda_c$.
%
 
\begin{figure}
  \hskip1.6in 
  \psfig{figure=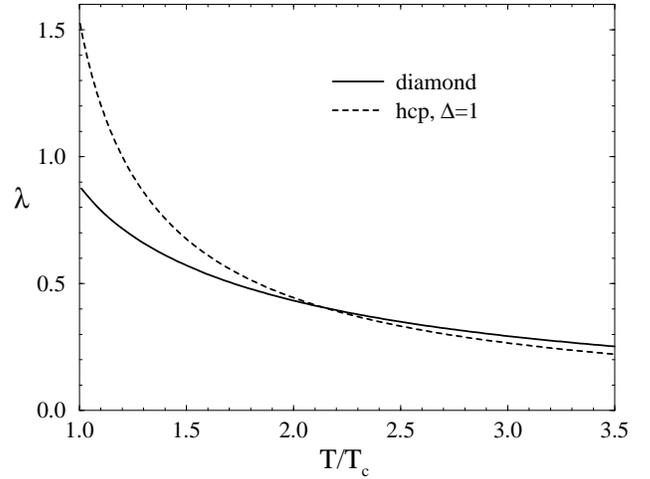,angle=-90.0,width=3.5in}
\caption{
\label{Fig-lambda}
The reaction field $\lambda$ for $T>T_c$ for the diamond lattice
and isotropic hexagonal close packed lattice ($\Delta=1$).
}
\end{figure}

Results for $\lambda(T)$ for the diamond lattice and
hcp lattice (at $\Delta=1$) are shown in Fig.\ \ref{Fig-lambda}.
The hcp lattice, which has higher coordination number,
also has the stronger reaction field at $T_c$.
As the temperature is increased, the reaction fields
diminish and become of comparable sizes for $T>2T_c$.
It is also expected that the slope relates to the
specific heat.\cite{XY} 
The graph then reasonably demonstrates a larger specific 
heat for hcp compared to diamond near $T_c$, in contrast 
to more similar specific heats at higher temperatures.

\section{Conclusions}
We have reviewed the standard Onsager reaction field approximation
for estimating $T_c$, and have shown how it can be extended to apply
to a Bravais lattice with a basis.
The bcc lattice was used as a test case because it can be calculated either
by the standard approach or the new method, when considered as sc with a
two-atom basis.
We used the new method to get $T_c$ for diamond and anisotropic hcp 
lattice systems, however, it certainly can be extended to more complex 
systems with a greater number of atoms per unit cell.
For the hcp lattice system, the ORF procedure used here was
found to be exactly equivalent to a Green's function (plus RPA) 
approach used by Adler.\cite{Adler81}
While it is an approximate method, it does give reasonable estimates
of $T_c$ and other quantities where other methods may be more cumbersome
or time-consuming to apply.
 
\medskip
{\sl Acknowledgments.}---Conversations with M.E. Gouvea,
J. Plascak, A.S.T. Pires and J. Kaplan are gratefully acknowledged.   
This work was partially supported by NSF/CNPq International 
Grant No.\ INT-9502781,  by a FAPEMIG Grant and by the Universidade 
Federal de Minas Gerais in Belo Horizonte, Brazil, where the work 
was completed.

\begin{references}
\bibitem{Onsager36} L. Onsager, J. Am. Chem. Soc. {\bf 58}, 1486 (1936).
\bibitem{Brout67} R. Brout and H. Thomas H., 
	Physics (Long Island City, N.Y.) {\bf 3}, 317 (1967).
\bibitem{spinglass} M. Cyrot, Phys. Rev. Lett. {\bf 43}, 173 (1979);
 	P. Nozi\`eres, J. Phys.(Paris) {\bf 43}, L-543 (1982).
\bibitem{electron} A. Georges and J.S. Yedidia, Phys. Rev. B {\bf 43}, 3475 (1991);
 	J.B. Stauton and B.L. Gyorffy, Phys. Rev. Lett. {\bf 69}, 371 (1992);
 	P. Kopietz, Phys. Rev. B {\bf 48}, 13789 (1993);
 	Y.H. Szczech, M.A. Tusch and D.E. Logan,
 	Phys. Rev. Lett. {\bf 74}, 2804 (1995).
\bibitem{hubbard} M. Cyrot and H. Kaga, Phys. Rev. Lett. {\bf 77}, 5134, (1996).
\bibitem{heisen} D.E. Logan, Y.H. Szczech and M.A. Tusch,
        Europhys. Lett. {\bf 30}, 307 (1995);
	M.P. Eastwood and D.E. Logan, Phys. Rev. B {\bf 52}, 9455 (1995).
\bibitem{XY} A.S.T. Pires, Solid State Comm. {\bf 98}, 933 (1996);
	M.\ E.\ Gouv\^ea and A.\ S.\ T.\ Pires, 
	Phys.\ Rev. B {\bf 54}, 14,907 (1996);
	M.E. Gouv\^ea, A.S.T. Pires and G.M. Wysin, 
	Phys. Rev. B {\bf 58}, 2399 (1998).
\bibitem{White83} R.M. White, {\it Quantum Theory of Magnetism}, 
	Sec.\ 4.1.3, Springer Series in Solid State Sciences, Vol.\ 32,
	Springer-Verlag (Berlin, Heidelberg, New York 1983).
\bibitem{Weiss48} P.\ R.\ Weiss,
	Phys.\ Rev.\ {\bf 74}, 1493 (1948).
\bibitem{Wysin99} G.M. Wysin and J. Kaplan, submitted to Phys. Rev. E.
\bibitem{He3} T.\ Lang, P.\ L.\ Moyland, D.\ A.\ Sergatskov,
	E.\ D.\ Adams and Y. Takano,  
	Phys.\ Rev.\ Lett.\ {\bf 77}, 322 (1996);
	H. Ishimoto, T.\ Okamoto, H.\ Fukuyama and  H.\ Akimoto, 
	J.\ Low Temp.\ Phys.\ {\bf 101}, 71 (1995);
	Y.\ Takano et al., Phys.\ Rev.\ Lett.\ {\bf 55}, 1490 (1985).
\bibitem{Fisher67} M.\ E.\ Fisher,
	Repts.\ Prog.\ Phys.\ {\bf 30} (pt.\ II), 615 (1967).
\bibitem{AM76} {\it Solid State Physics}, N.\ Ashcroft and N.\ Mermim,
	Holt, Reinhart and Winston (1976).
\bibitem{Chen93} K.\ Chen, A.\ M.\ Ferrenberg and D.\ P.\ Landau,
	Phys.\ Rev.\ B {\bf 48}, 3249 (1993). 
\bibitem{Wu82} F.\ Y.\ Wu, Rev.\ Mod. Phys. {\bf 54}, 235 (1982).
\bibitem{Domany84} E.\ Domany, Phys.\ Rev.\ Lett.\ {\bf 52}, 871 (1984).
\bibitem{Domany+85} E.\ Domany and J.\ E.\ Gubernatis, 
	Phys.\ Rev.\ B {\bf 32}, 3354 (1985).
\bibitem{Auerbach+88} D.\ Auerbach, E.\ Domany and J.\ E.\ Gubernatis,
	Phys.\ Rev.\ B {\bf 37}, 1719 (1988).
\bibitem{Adler81} J. Adler, Physica {\bf 107B}, 207 (1981).
\bibitem{AdlerNote} Our hcp ORF results agree exactly with the
	results of Ref. \protect\cite{Adler81} when we use
	the relation $n k_B T_c = 6J(1+\Delta)/F(-1)$, with $n=3$,
	rather than $n=2$ as printed there.
\bibitem{Domb+57} C.\ Domb and M.\ F.\ Sykes, Proc.\ Phys.\ Soc.\ 
	Lond.\ {\bf A 70}, 326 (1957).
\end{references}
\end{document}